\documentclass{ifacconf}

\usepackage{graphicx}      % include this line if your document contains figures
\usepackage{natbib}        % required for bibliography
\usepackage{amssymb}
\usepackage{amsmath}
\begin{document}
\begin{frontmatter}

\title{Knowledge extraction, modeling and formalization: EEG case study} 
%\title{Mining of Characteristic Properties of Sleep Spindles in Electroencephalographic Datasets with Formal Concept Analysis}

\author[1]{Dmitry Morozov} 
\author[1]{Mario Lezoche} 
\author[1]{Herv\'e Panetto}

\address[1]{Research Center for Automatic Control of Nancy (CRAN), Universit\'e de Lorraine, UMR 7039, Boulevard des
Aiguillettes B.P.70239, 54506 Vandoeuvre-l\`es-Nancy, France.\\ e-mail: \{dmitrii.morozov, mario.lezoche, herve.panetto\}@univ-lorraine.fr}

\begin{abstract} 
Formal Concept Analysis (FCA) is a well-established method for data analysis which finds many applications in data mining.  Its extension on complex data representation formats brought a wave of new applications to the problems such as gene expression mining \cite{a15}, prediction of toxicity of chemical compounds \cite{a19} or clustering of sequences in process event logs \cite{a18} . Insipired from this work our research inherits their model and designs an experiment for mining electroencephalographic recordings for patterns of sleep spindles. The contribution of this paper lies in the specification of desritizition procedure and the architecture of FCA experiment. We also provide some reflection on the related research papers.
\end{abstract}


\end{frontmatter}
%===============================================================================

\section{INTRODUCTION}
Neuroscience which is primarily a biological science in fact offers wide range of problems to different scientific communities and stimulates  interdisciplinary research across various domains. In this way mathematical graph theory serves as an instrument to analyze and build models of connectivity of brain regions (\cite{a11, a16}). Advances in Machine learning methods allow to construct accurate predictions for epilepsy seizures (\cite{a12}) or gain insights into the nature of sleep (\cite{a13}).

Formal concept analysis (\cite{a1}) which was initially borned from the algebraic lattice theory to the moment has took rather applicational direction. It provides methods to structure and explore the domain of interest, and discover new knowledge. 

This study takes a look at application of Formal Concept Analysis (FCA) for segregation and differentiating sleep spindles, a kind of waves occurring on an EEG during sleep. They are characterized by short burst of high frequency brain activity and primarily serve as an indicator of a stage 2 sleep.

The rest of the paper is organized as follows. The second section introduces the methods of FCA and its extensions that lay the basis of our approach. The third section dedicated to the details and the design of conducted experiment. In the forth section we discuss related research; and the final section concludes the paper with a summary.

\section{METHODS}
\subsection{Formal Concept Analysis (FCA)}
FCA \cite{a1} is at its core a mathematical formalism which with time has developed and been extended with many theoretical and applied studies. Starting with a set of objects and a set of attributes FCA finds generalizations of the descriptions for arbitrary subset of objects.

Let $G$ and $M$ be sets, called the set of objects and attributes, respectively, and let $I$ be a relation $I\subseteq G\times M$: for $g\in G, m\in M, gIm$ holds iff the object $g$ has the attribute $m$. The triple $\mathrm{K}=(G,\ M,\ I)$ is called $\mathrm{a}$ {\it formal context}.

If $A\subseteq G, B\subseteq M$ are arbitrary subsets, then the {\it Galois connection} is given by the following {\it derivation operators}:

$A'=\{m\in M|gIm$ for all $g\in A\},$

$B'=\{g\in G|gIm$ for all $m\in B\}.$

The pair $(A,\ B)$ , where $A\subseteq G, B\subseteq M, A'=B$, and $B'=A$ is called a {\it formal concept of the context} $K$. $A$ is called the {\it extent}  and $B$ the {\it intent} of the formal concept $(A,\ B)$. From the properties of the derivation operators it follows that the conditions $A'=B$,  $B'=A$ can be represented in more simple way $A''=A$, or equivalently $B''=B$. This reformulated form signifies that a formal concept is such a pair of sets that either of them is {\it closed} under derivation operator ($\cdot$)'.

The concepts, ordered by $(A_{1},\ B_{1})\geq(A_{2},\ B_{2}) \Leftrightarrow A_{1}\supseteq A_{2}$ form a complete lattice, called {\it the concept lattice} $L(G,\ M,\ I)$ .

\subsection{Pattern Structures}
Pattern Structures are generalization of FCA that extends its theory beyond binary formal contexts on other formats of description (e.g. sequences or graphs).

{\it A pattern structure} $\mathbb{P}$ { is a triple} $(G,\ (D,\ \sqcap),\ \delta)$ , { where} $G, D$ { are sets, called the set of objects and the set of descriptions, and} $\delta$ : $G\rightarrow D$ { maps an object to a description. Respectively}, $(D,\ \sqcap)$ { is a meet-semilattice on} $D$ { w.r.t}. $\sqcap$, { called similarity operation such that} $\delta(G):=\{\delta(g) \ | \ g\in G\}$ { generates a complete subsemilattice} $(D_{\delta},\ \sqcap)$ { of} $(D,\ \sqcap)$ .

It can be noticed that FCA fits to the definition pattern structures. The set of objects $G$ is preserved, the semilattice of descriptions is $(\wp(M),\ \cap)$ , ( $\wp(M)$ denotes the powerset of the set of attributes $M$), a description is a subset of attributes and $\cap$ is the set-theoretic intersection. $\delta$ : $G\rightarrow\wp(M)$ is given by $\delta(g)=\{m\in M|(g,\ m)\in I\}.$

The Galois connection for a pattern structure $(G,\ (D,\ \sqcap),\ \delta)$ , relating sets of objects and descriptions, is defined as follows:

$A^{\mathrm{\diamond}}:=\sqcap_{g\in A}\delta(g)$ for $A\subseteq G$

$d^{\mathrm{\diamond}}:=\{g\in G | d\sqsubseteq\delta(g)\},$ for $d\in D$

Given a subset of objects $A, A^{\mathrm{\diamond}}$ returns the description which is common to all objects in $A$. Given a description $d, d^{\mathrm{\diamond}}$ is the set of all objects whose description subsumes $d$. The natural partial order (or subsumption order between descriptions) $\sqsubseteq$ on $D$ is defined w.r.t. the similarity operation $\sqcap:c\sqsubseteq d\Leftrightarrow c\sqcap d=c$ (in this case we say that $c$ is subsumed by $d$). In the case of standard FCA the natural partial order corresponds to the set-theoretical inclusion order, i.e., for two sets of attributes $x$ and $y$ $x\sqsubseteq y\Leftrightarrow x\subseteq y.$

{\it A pattern concept of a pattern structure} $(G,\ (D,\ \sqcap),\ \delta)$ { is a pair} $(A,\ d)$ , { where} $A\subseteq G$ { and} $d\in D$ { such that} $A^{\mathrm{o}}=d$ { and} $ d^{\mathrm{o}}=A,\cdot A$ { is called the pattern extent and} $d$ { is called the pattern intent}.

As in standard FCA, a pattern concept corresponds to the maximal set of objects $A$ whose description subsumes the description $d$, where $d$ is the maximal common description of objects in $A$. The set of all pattern concepts is partially ordered w.r.t. inclusion of extents or, dually, w.r.t. subsumption of pattern intents within a concept lattice, these two antiisomorphic orders making a lattice, called pattern lattice.

\subsection{Example of a Pattern Structure}

The most simple extension of binary context can be made by introduction of numbers. And Interval pattern structure provides a natural way to organize descriptions of subsets of objects into a lattice for such a case. Interval pattern structures were introduced by \cite{a15} where they were used for gene expression analysis. 

\begin{table}[h]
\begin{center}
\caption{Example: an interval context} 
\label{IPScontext}
\begin{tabular}{c|cc}
\ & $m_1$ & $m_2$\\
\hline
\noalign{\vskip 1mm}
$g_1$ & [1, 1] & [1, 1] \\
$g_2$ & [2, 2] & [2, 2] \\
$g_3$ & [3, 3] & [2, 2] \\
\end{tabular}

\end{center}
\end{table}

Table \ref{IPScontext} shows an example of an interval context where three objects are described by two attributes. In such pattern structure values of attributes are represented by intervals, but the meaning is not probabilistic. Intervals represent a way to generalize underlying values and to ascend up a level of abstraction. 

If we have two objects, then a numerical attribute can have all values from the interval of this attribute in the first object and from the interval of this attribute of the second object. Consequently, the similarity between two intervals can be defined as a convex hull of the intervals, i.e. $[a,\displaystyle \ b]\sqcap[c,\ d]=[\min(a,\ c)$ , $\displaystyle \max(b, d)]$ Then, given two tuples of intervals, the similarity between these tuples is computed as a component-wise similarity between intervals.

In this example, we have the pattern structure $(G,\ (D,\ \sqcap),\ \delta)$ , where $G= \{g_{1},\ g_{2},\ g_{3}\}$, the set $D$ is the set of all possible interval pairs with the similarity operation described above, and $\delta$ is given by the context in Table \ref{IPScontext}  i.e., $\delta(g_{1})=\langle[1$, 1 $]$; $[$1, $ 1]\rangle$ and $\delta(g_{1})\sqcap\delta(g_{2})=\langle[1, 2 ]; [1, 2]\rangle$

\begin{figure}
\begin{center}
\includegraphics[width=8.4cm]{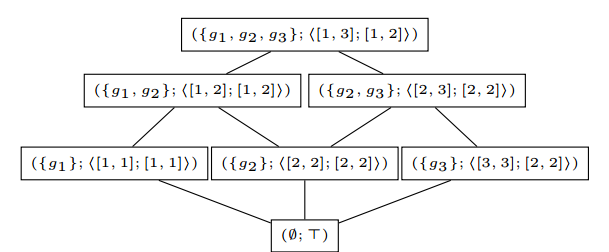}  
\caption{The diagram of the lattice of for the Interval context given in table \ref{IPScontext}} 
\label{IPSexample}
\end{center}
\end{figure}

Figure \ref{IPSexample} shows the resulting pattern lattice organizing all the pattern concepts of the Table \ref{IPScontext} into an hierarchy.

\subsection{Index of stability}
Stability indexes were introduced by \cite{a3}. One distinguishes intentional and extensional stability. The first one allows estimating the strength of dependence of an intent on each object of the respective extent. Extensional stability is defined in a dual way.
$$
Stab_{i}(A,\ B)=\frac{|\{C\subseteq A|C'=B\}|}{2^{|A|}}
$$

Stability of a formal concept may be interpreted as probability of retaining its intent after removing some objects from the extent (assuming that all the subsets of the extent are equally probable).

Calculation of stability is been shown to be intractable (\cite{a3}) implying high computation demands of the problem. Although there have been suggested a number of methods for its approximate calculation. According to \cite{a14} for large contexts the stability is close to 1 and it appears to be more practical to use logarithmic scale of stability inducing the same ranking as stability :

{\it LStab} $(c)=-log_{2}$( $1$ --{\it Stab} $(c)$ )

In this work they also suggest the following bounds of stability for its approximate and fast calculation:
$$\Delta_{\min}(c)-log_{2}(|M|) \leq$$
$$\leq  -log_{2}\sum_{d\in DD(c)}2^{-\Delta(c,d)} \leq LStab(c)\leq\Delta_{\min}(c)$$

where $\displaystyle \Delta_{\min}(c)=\min_{d\in DD(c)}\Delta(c,\ d)$ , $DD(c)$ is a set of all direct descendants of $c$ in the lattice and $\Delta(c,\ d)$ is the size of the set-difference between extents of formal concepts $c$ and $d.$

\section{EXPERIMENT}
\subsection{Data}
The main goal behind the experiment was to explore different characteristic patterns of sleep spindles and to select the most significant and descriptive. Although ultimate significance of found patterns can only be judged by medical experts we use established metrics of Data Mining as an intermediate evaluation. These metrics are support and stability index.

For our experiment we use a data-set of EEG recordings provided by Neurology ward of Central Hospital of Nancy. Data-set represents set of signals captured from 19 electrodes placed on a hospitalized patient's scalp according to 10 -- 20 system . Overall duration of the signal is roughly 6 minutes and a meta-file provides timings of the sleep spindles. 

\begin{figure}
\begin{center}
\includegraphics[width=8.4cm]{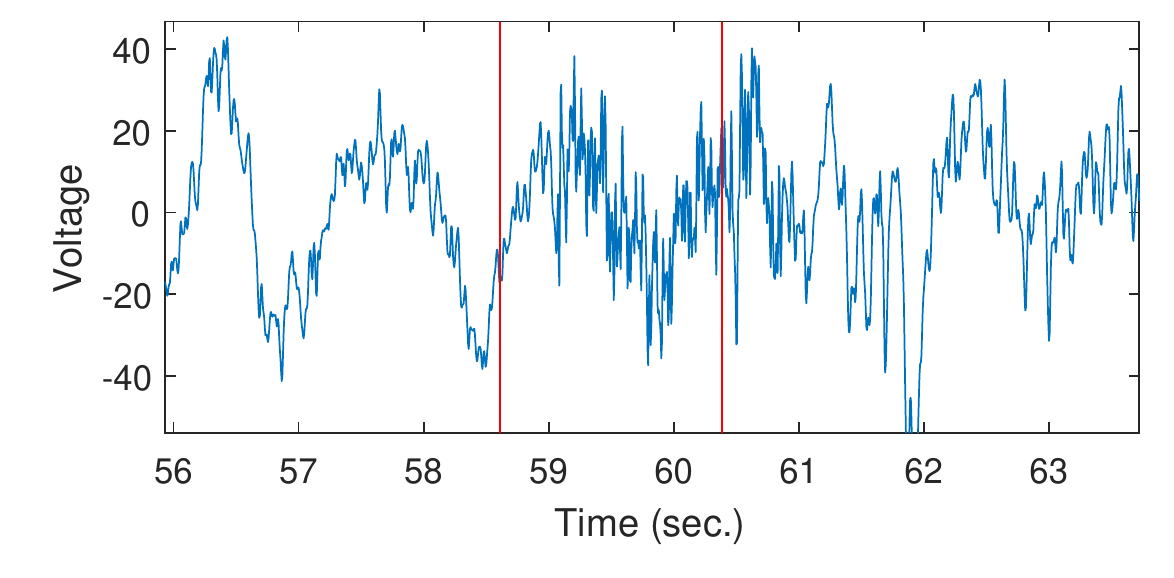}  
\caption{Example: the curve bounded by red lines corresponds to a spindle from the electrode F4 of one of the EEG data-sets} 
\label{fig:spindle}
\end{center}
\end{figure}

An example of a spindle from one of the patents' EEG is represented on the Figure \ref{fig:spindle}. The curve eclosed between two red vertical lines  stands out from the rest of the signal with its elevated frequency. What are the characteristic properties that the spindle posses? Are there any other types of spindles? We aim to answer these questions with analysis of EEG with an FCA approach .

\subsection{Formation of formal context}
\begin{figure}
\begin{center}
\includegraphics[width=8.4cm]{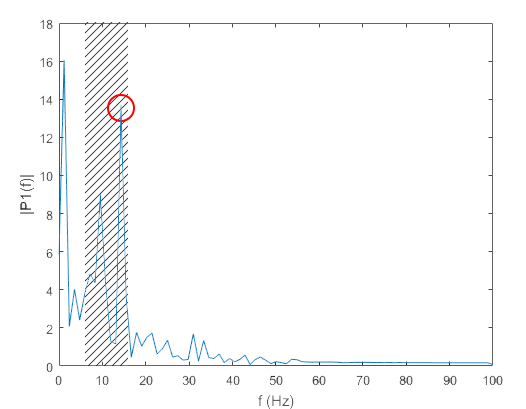}  
\caption{Extraction of dominant frequency as a maximum of Furier transform of the spindle's signal in the characteristic interval of 6 -- 14 Hz} 
\label{fig:dominant_freq}
\end{center}
\end{figure}

As FCA requires description of the domain in discrete way the fundamental part of setup of an experiment is transition from continuous signals to particular descriptions of their properties. We achieve this by calculating derivative characteristics of signals of sleep spindles. In this way formal context has sleep spindles as objects, and they are described with properties of their signal. The following is the list of standard properties used by EEG-analysts for detection of spindles that we in our turn adopted for our experiment:

\begin{itemize}
\item $\langle A \rangle$ average amplitude 
\item $A_{max}$ maximum of absolute value of amplitude
\item $\langle f \rangle$  average frequency
\item  $\mathbf{f}$ dominant frequency (see Fig. \ref{fig:dominant_freq})
\item $\langle A \rangle / \mathbf{f}$ ratio of average amplitude to dominant frequency
\item $P_{[0.5 - 2.5]}$ \dots average bandpower for frequency bands [0,5 -- 2,5], [2.5 -- 4.5], \dots [28,5 -- 30,5]
\end{itemize}

\begin{table}
\caption{First 6 columns of the formal context. Each spindle (on the raw) is represented by the properties of its signal} 
\label{table:context}
\begin{center}
\begin{tabular}{rrrrrr}
\hline
\noalign{\vskip 1mm}

$\langle A \rangle$ & $A_{max}$ & $\langle f \rangle$ & $\mathbf{f}$ & $\langle A \rangle / \mathbf{f}$ & $P_{[0.5 - 2.5]}$ \\
\hline
\noalign{\vskip 2mm}
14,76 & 43,39 & 7,06 & 11,56 & 1,28 & 111,74 \\
13,57 & 49,64 & 7,11 & 7,69 & 1,76 & 162,25\\
16,61 & 41,03 & 7,28 & 9,52 & 1,74 & 122,01 \\
14,66 & 54,78 & 5,07 & 8,24 & 1,78 & 270,57 \\
15,95 & 46,12 & 4,95 & 11,34 & 1,41 & 142,78\\
16,58 & 54,96 & 3,82 & 11,67 & 1,42 & 290,43\\
13,16 & 45,85 & 5,36 & 11,48 & 1,15 & 134,28\\
15,62 & 51,31 & 4,89 & 10,11 & 1,54 & 238,48\\
12,21 & 31,85 & 4,86 & 11,81 & 1,03 & 86,07\\
12,55 & 41,17 & 7,79 & 8,00 & 1,57 & 129,01\\
14,35 & 65,78 & 4,25 & 10,92 & 1,31 & 385,95\\
17,66 & 48,67 & 4,87 & 9,13 & 1,93 & 449,57\\
14,94 & 39,12 & 5,47 & 9,73 & 1,53 & 281,13\\
20,77 & 74,99 & 3,66 & 9,66 & 2,15 & 623,86\\
17,83 & 64,39 & 5,64 & 12,11 & 1,47 & 128,97\\
16,00 & 43,11 & 3,89 & 11,89 & 1,35 & 329,20\\
15,40 & 60,87 & 6,83 & 12,14 & 1,27 & 234,18\\
9,47 & 29,63 & 7,81 & 7,56 & 1,25 & 39,45\\
16,11 & 42,39 & 3,46 & 10,45 & 1,54 & 271,81\\
18,42 & 67,76 & 3,47 & 11,76 & 1,57 & 147,11\\

 \hline

\end{tabular}
\end{center}

\end{table}
For calculation of dominant and average frequency we  used Furier transform and MATLAB meanfreq correspondingly. Analysis of different frequency bands was conducted with the help of MATLAB bandpower function. The result of calculation of specified properties is a formal context that serves as an input for FCA algorithms. 

Table \ref{table:context} depicts a part of the final context constituting the subject of our experiment. On top of that, the whole context has additional columns for the bandpower attributes, which are not included here because of big size of the resulting table.

\subsection{Scheme of the experiment}

Table \ref{table:scheme} lists the steps of the execution sequence of the experiment. First stage is preprocessing of the data-set. It includes extraction of signals of the spindles from the entire input EEG data-set. It means that each spindle present in the data-set will be given a set of cut-out signals from all the electrodes describing only particular moments in time when that spindle occured.

 Then follows application of signal treatment procedures for formation of a context according to the details given in the previous section. Attribute selection which is a standard Data Mining methodology is utilized to reduce dimensionality and decrease further computational costs of treating the data. It uses information gain metric and correlation between attributes to select a subset of the most relevant attributes of the dataset.

The second stage of the experiment which marks engagement of FCA methodology is construction of Formal concept diagram. Two approaches of producing and processing FCA diagram find their place in practice: one that require keeping the whole diagram in the memory and the other which processes the results on-line and discards concepts as they go. We will be using the first more demanding approach which on the other hand enables us to colculate stability index. For construction of a lattice we use the algorithm suggested by \cite{a17}.

The third stage is dedicated to traversal of the lattice and calculation of stability for each of its concepts.We use boundaries described in the section 2.4 for its fast approximation.  After it, filtering is applied to the whole lattice, stage 4, giving in the result a list of patterns satisfying the creteria.

Implementation-vise, preprocessing stage is carried out with help of ready-to-use MATLAB procedures, putting  all the necessary work of data-format manipulations on python scripts. Formal concept analysis algorithms for construction of concept lattice and calculation of stability use C++ implementation.

\begin{table}[h!]
\large
\caption{Mining of stable frequent patterns of sleep spindles in EEG recordings} 
\label{table:scheme}
\begin{center}
\begin{tabular}{l}
 \hline
\noalign{\vskip 2mm}
{\it Input: EEG recordings of brain activity}, \\ {\it meta file providing spindle start and end timings}\\
1. Preprocessing of raw EEG data: \\
	\qquad Extraction of signals of the spindles \\
	\qquad Calculation of the characteristics\\
	\qquad Formation of a formal context\\
	\qquad Attribute selection\\
2. Construction of the Formal concept diagram\\
3. Calculation of stability index of the concepts\\
4. Filtering of patterns by stability and support\\

{\it Output: patterns meeting specified constraints }\\
\noalign{\vskip 2mm}
 \hline
\end{tabular}
\end{center}

\end{table}

\section{RELATED STUDIES}
Our research falls in line with the recent publications (\cite{a6,a7}) of practical applications of FCA to real data. Qualitative assessment of the results (\cite{a8,a9,a10}) shows that FCA methods are not weaker than other Data Mining techniques, but represent an attractive alternative. Nevertheless it can seem that FCA beers with itself unnecessary power, which also affects the amount of resources it uses compared to lighter statistical methods. 

In the literature articles on application of FCA to neuroimaging have already appeared. \cite{a4} utilizes FCA to discover hierarchy in the neural codes analyzing big set of fMRI data. Although the size of  the impact of data filtering and approximation preceding utilization of FCA is not quite clear. There is no convenient method to assess the cost of putting the data on a coarse scale when the end result, and so the effect, cannot be measured in any way.

Another study by \cite{a5} uses FCA methodology to separate neural patterns of white noise, in the spike train data where the signal was captured directly from implanted electrodes. Although they approve the hypothesis that stability measure is suitable to filter out noise, the design of the experiment is not clearly leads to the conclusion. In some settings it is admissible to train a classifier on negative sample, so that in the end they can be differed from positive examples. Although this approach is highly prone to overfitting. 

But the main drawback is that stability as a distinctive criterion is not well suitable for that purpose, because it is not preserved when moving from training to test data-set. The same formal concepts on test and training data-sets will have different values of stability. Which puts under the question the applicability of traditional training and classification model for this case.

\section{CONCLUSION}
In this paper we presented a preliminary study of application of FCA to EEG recordings. We specified data transformation protocol and a method to filter patterns by their significance.

Current vector of work is directed at resolving the issue of scalability of FCA techniques and at an additional search criteria that could improve how good found patterns can be interpreted.

From interoperability perspective, future research will be dedicated to application of FCA approach to study interrelation between the insights provided by EEG and fMRI techniques.

\nocite{*}
\bibliography{longnames}

\begin{thebibliography}{19}
\providecommand{\natexlab}[1]{#1}
\providecommand{\url}[1]{\texttt{#1}}
\providecommand{\urlprefix}{URL }
\expandafter\ifx\csname urlstyle\endcsname\relax
  \providecommand{\doi}[1]{doi:\discretionary{}{}{}#1}\else
  \providecommand{\doi}{doi:\discretionary{}{}{}\begingroup
  \urlstyle{rm}\Url}\fi

\bibitem[{Acharya et~al.(2005)Acharya, Fausta, Kannathala, Chuaa, and
  Laxminarayanb}]{a13}
Acharya, R., Fausta, O., Kannathala, N., Chuaa, T., and Laxminarayanb, S.
  (2005).
\newblock Non-linear analysis of eeg signals at various sleep stages.
\newblock \emph{Computer Methods and Programs in Biomedicine}, 80, 37—--45.

\bibitem[{Buzmakov et~al.(2014)Buzmakov, Kuznetsov, and Napoli}]{a14}
Buzmakov, A., Kuznetsov, S., and Napoli, A. (2014).
\newblock Scalable estimates of concept stability.
\newblock In \emph{Glodeanu, C., Kaytoue, M., Sacarea, C. (eds.) Formal Concept
  Analysis}, volume 8478 of \emph{Lecture Notes in Computer Science},
  157--–172. Springer.

\bibitem[{Chu et~al.(2012)Chu, Kramer, Jay~Pathmanathan, Westover, Wizon, and
  Cash}]{a11}
Chu, C.J., Kramer, M.A., Jay~Pathmanathan, M.T.B., Westover, M.B., Wizon, L.,
  and Cash, S.S. (2012).
\newblock Emergence of stable functional networks in long-term human
  electroencephalography.
\newblock \emph{Journal of Neuroscience}, 32(8), 2703--2713.

\bibitem[{Endres et~al.(2012)Endres, Adam, Giese, and Noppeney}]{a4}
Endres, D., Adam, R., Giese, M.A., and Noppeney, U. (2012).
\newblock Understanding the semantic structure of human fmri brain recordings
  with formal concept analysis.
\newblock In \emph{10th International Conference, ICFCA 2012, Leuven, Belgium,
  May 7-10, 2012. Proceedings}, volume 7278 of \emph{Lecture Notes in Computer
  Science}, 96--111. Springer.

\bibitem[{Ganter and Kuznetsov(2001)}]{a2}
Ganter, B. and Kuznetsov, S. (2001).
\newblock Pattern structures and their projections.
\newblock In \emph{(ICCS 2001)}, volume 2120 of \emph{Lecture Notes in
  Artificial Intelligence}, 129--142. Springer.

\bibitem[{Ganter and Wille(1997)}]{a1}
Ganter, B. and Wille, R. (1997).
\newblock \emph{Formal Concept Analysis: Mathematical Foundations}.
\newblock Springer, NJ, USA.

\bibitem[{Giusti et~al.(2015)Giusti, Pastalkova, Curto, , and Itskov}]{a16}
Giusti, C., Pastalkova, E., Curto, C., , and Itskov, V. (2015).
\newblock Clique topology reveals intrinsic geometric structure in neural
  correlations.
\newblock \emph{Proceedings of the National Academy of Sciences of the United
  States of America}, 112 no. 44, 13455–--13460.

\bibitem[{Kashnitsky and Ignatov(2014)}]{a9}
Kashnitsky, Y. and Ignatov, D.I. (2014).
\newblock Can fca-based recommender system suggest a proper classifier?
\newblock In \emph{Proceedings of the International Workshop "What can FCA do
  for Artificial Intelligence?" (FCA4AI at ECAI 2014)}, volume 1257, 17--26.
  CEUR Workshop Proceedings.

\bibitem[{Kashnitsky and Kuznetsov(2016)}]{a10}
Kashnitsky, Y. and Kuznetsov, S. (2016).
\newblock Interval concept lattice as a classifier ensemble.
\newblock In \emph{Proceedings of the International Workshop "What can FCA do
  for Artificial Intelligence?" (FCA4AI at ECAI 2016)}, volume 1703. CEUR
  Workshop Proceedings.

\bibitem[{Kaytoue et~al.(2011)Kaytoue, Kuznetsov, Napoli, , and
  Duplessis}]{a15}
Kaytoue, M., Kuznetsov, S.O., Napoli, A., , and Duplessis, S. (2011).
\newblock Mining gene expression data with pattern structures in formal concept
  analysis.
\newblock \emph{Inf. Sci. (Ny)}, 181(10), 1989 –-- 2001.

\bibitem[{Korepanova and Kuznetsov(2016)}]{a7}
Korepanova, N.V. and Kuznetsov, S.O. (2016).
\newblock Pattern structures for treatment optimization.
\newblock In \emph{CLA 2016: Proceedings of the Thirteenth International
  Conference on Concept Lattices and Their Applications}, volume 1624,
  217--230. CEUR Workshop Proceedings.

\bibitem[{Kourie et~al.(2009)Kourie, Obiedkov, Watson, and van~der Merwe}]{a17}
Kourie, D.G., Obiedkov, S., Watson, B.W., and van~der Merwe, D. (2009).
\newblock An incremental algorithm to construct a lattice of set intersections.
\newblock \emph{Science of Computer Programming}, 74 No. 3, 128--142.

\bibitem[{Kuznetsov(2007)}]{a3}
Kuznetsov, S. (2007).
\newblock On stability of a formal concept.
\newblock \emph{Annals of Mathematics and Artificial Intelligence}, 49,
  101--115.

\bibitem[{Kuznetsov and Makhalova(2015)}]{a8}
Kuznetsov, S. and Makhalova, T. (2015).
\newblock Concept interestingness measures: a comparative study.
\newblock In \emph{Proceedings of the Twelfth International Conference on
  Concept Lattices and Their Applications Clermont-Ferrand, France, October
  13-16}, volume 1466, 59--72. CEUR Workshop Proceedings.

\bibitem[{Kuznetsov and Samokhin(2005)}]{a19}
Kuznetsov, S. and Samokhin, M. (2005).
\newblock Learning closed sets of labeled graphs for chemical applications.
\newblock In \emph{Proc. 15th Conference on Inductive Logic Programming (ILP
  2005)}, volume 3625 of \emph{Lecture Notes in Artificial Intelligence},
  190--208. Springer.

\bibitem[{Masyutin and Kuznetsov(2016)}]{a6}
Masyutin, A. and Kuznetsov, S. (2016).
\newblock Continuous target variable prediction with augmented interval pattern
  structures: Lazy algorithm.
\newblock In \emph{CLA 2016: Proceedings of the Thirteenth International
  Conference on Concept Lattices and Their Applications}, volume 1624,
  273--284. CEUR Workshop Proceedings.

\bibitem[{Mirowski et~al.(2009)Mirowski, Madhavan, LeCun, and Kuzniecky}]{a12}
Mirowski, P., Madhavan, D., LeCun, Y., and Kuzniecky, R. (2009).
\newblock Classification of patterns of eeg synchronization for seizure
  prediction.
\newblock \emph{Clinical Neurophysiology}, 120, 1927–--1940.

\bibitem[{Morozov et~al.(2015)Morozov, Kuznetsov, Lezoche, and Panetto}]{a18}
Morozov, D., Kuznetsov, S., Lezoche, M., and Panetto, H. (2015).
\newblock Formal methods for process knowledge extraction.

\bibitem[{Yegenoglu et~al.(2016)Yegenoglu, Quaglio, Torre, Grün, and
  Endres}]{a5}
Yegenoglu, A., Quaglio, P., Torre, E., Grün, S., and Endres, D. (2016).
\newblock Exploring the usefulness of formal concept analysis for robust
  detection of spatio-temporal spike patterns in massively parallel spike
  trains.
\newblock In \emph{22nd International Conference on Conceptual Structures, ICCS
  2016, Annecy, France, July 5-7, 2016, Proceedings}, volume 9717 of
  \emph{Lecture Notes in Computer Science}, 3--16. Springer.

\end{thebibliography}
                                                  
\end{document}